\documentclass[nofootinbib, aps,twocolumn,showpacs,preprintnumbers,amssymb,superscriptaddress,10pt]{revtex4-1}
\pdfoutput=1

\usepackage{amsmath,amssymb,amsfonts,graphicx,epsfig,todonotes}
\usepackage[noconfig]{refstyle}
\usepackage[hyperfootnotes=true, linktocpage=true, citecolor=blue, linkcolor=blue, urlcolor=Maroon, filecolor=Maroon]{hyperref}

\usepackage{multirow}
\usepackage{adjustbox}
\usepackage{subcaption}

\let\eqref=\relax
\newref{eq}{name={},Name={Eq.~},names={eqs.~},Names={Eqs.~},rngtxt={-},refcmd=(\ref{#1})}
\newref{tab}{name={Table~},Name={Table~},names={tables~},Names={Tables~}}
\newref{sec}{name={Section~},Name={Section~},names={sections~},Names={Sections~}}
\newref{fig}{name={Figure~},Name={Figure~},names={figures~},Names={Figures~}}
\numberwithin{equation}{section}

\newcommand{\comment}[1]{}

\newcommand{\IZ}{\mathbb{Z}}

\begin{document}
\title{Machine Learning Lie Structures \& Applications to Physics}

\author{Heng-Yu Chen}
\affiliation{Department of Physics,National Taiwan University, Taipei 10617, Taiwan}
\email[]{heng.yu.chen@phys.ntu.edu.tw}

\author{Yang-Hui He}
\affiliation{Department of Mathematics, City, University of London, London EC1V0HB, UK}
\email[]{Suvajit.Majumder@city.ac.uk}
\affiliation{Merton College, University of Oxford, OX1 4JD, UK}
\affiliation{School of Physics, NanKai University, Tianjin, 300071, P.R. China}
\email[]{hey@maths.ox.ac.uk}

\author{Shailesh Lal}
\affiliation{Faculdade de Ciencias, Universidade do Porto,  687 Rua do Campo Alegre, Porto 4169-007, Portugal}
\email[]{shailesh.hri@gmail.com}

\author{Suvajit Majumder}
\affiliation{Department of Mathematics, City, University of London, London EC1V0HB, UK}

\begin{abstract}
Classical and exceptional Lie algebras and their
representations are among the most 
important tools in the analysis of symmetry in
physical systems. In this letter we show how the computation
of tensor products and branching rules of irreducible representations
are machine-learnable, and can achieve relative speed-ups of
orders of magnitude in comparison to the non-ML algorithms.
\end{abstract}


\maketitle

\section{Introduction \& Summary}
Lie algebras are an integral part of modern mathematical physics.
Their representation theory governs every field of physics from the fundamental structure of particles to the states of a quantum computer.
Traditionally, an indispensable tool to the high energy physicist is the extensive tables of \cite{Slansky:1981yr}.
More contemporary usage, with the advent of computing power of the ordinary laptop, have relied on the likes of highly convenient software such as ``LieART'' \cite{Feger:2019tvk}. Such computer algebra methods, especially in conjunction with the familiarity of the Wolfram programming language to the theoretical physicists, are clearly destined to play a helpful r\^ole.

In parallel, a recent programme of applying the techniques from machine-learning (ML) and data science to study various mathematical formulae and conjectures had been proposed \cite{He:2017aed,He:2017set,He:2018jtw}.
Indeed, while the initial studies were inspired by and brought to string theory in timely and independent works in  \cite{He:2017aed,He:2017set,Krefl:2017yox,Ruehle:2017mzq,Carifio:2017bov},
experimentation of whether standard techniques in neural regressors and classifiers could be carried over to study diverse problems have taken a life of its own. 
These have ranged from finding bundle cohomology on varieties \cite{Ruehle:2017mzq,Brodie:2019dfx,Larfors:2020ugo}, to distinguishing elliptic fibrations \cite{He:2019vsj} and invariants of Calabi-Yau threefolds \cite{Bull:2018uow}, to machine-learning the Donaldson algorithm for numerical Calabi-Yau metrics \cite{Ashmore:2019wzb},
to the algebraic structures of discrete groups and rings \cite{He:2019nzx}, to the BSD conjecture \& Langlands programme in number theory \cite{Alessandretti:2019jbs,He:2020eva,He:2020kzg}, to quiver gauge theories and cluster algebras \cite{Bao:2020nbi}, 
to patterns in particle masses \cite{Gal:2020dyc}, to knot invariants \cite{knots},
to statistical predictions and model-building in string theory \cite{Deen:2020dlf,Halverson:2019tkf,Halverson:2020opj}, to classifying combinatorial properties of finite graphs \cite{He:2020fdg} etc.
Moreover, the very structures of quantum field theory and holography \cite{Akutagawa:2020yeo,Koch:2019fxy,Halverson:2020trp} have also been proposed to be closely related to suitable neural networks.

In this letter, we continue this exciting programme and apply machine learning techniques to another indispensable concept for physicists, namely the ubiquitous continuous symmetries as encoded by Lie groups/algebras. Physicists have long used them to classify from the phases of matters to the spectrum of elementary particles.
As listed earlier, machine learning techniques have provided us with a powerful new approach towards various classification problems of physical interests\footnote{
There have been other interesting works on detecting physical symmetries using machine-learning \cite{Krippendorf:2020gny,Chen:2020dxg}.}.
Here we would like to ask whether the essential structures of Lie group can also be learned by machine.
Specifically, by this we mean whether neural nework (NN) classifiers and regressors can, after having seen enough samples of typical calculations such as tensor decomposition or branching rules -- both known to be heavily computationally expensive, as we will shortly see -- predict the result more efficiently.

As a comparison, let us also mention a somewhat surprising result from \cite{He:2019nzx}, where some fundamental structures of algebra, viz., certain properties of finite groups and finite rings, seem to be machine-learnable.
Difficult problems in representation theory such as recognizing whether a finite group is simple or not by ``looking'' at the Cayley multiplication table, or whether random permutation matrices (Sudoku) possess group structure, etc., can be classified by a support vector machine very efficiently without recourse to the likes of Sylow theorems which are computationally expensive.

In this letter we are motivated by the question of whether and how much one can machine-learn the essential information about classical, and exceptional Lie algebras as tabulated in standard texts such as Slansky \cite{Slansky:1981yr}.
Specifically, we address the two fundamental problems in the representation theory of Lie algebras that is crucial to physics -- the 
tensor product decomposition and the branching rules to a sub-algebra -- and show that these salient structures are machine
learnable. 

In particular we show that a relatively simple forward-feeding neural network can predict to high accuracy and confidence, 
the number of irreducible representations (``irreps'') that appear in a tensor
product decomposition, which we refer to as the \textit{length} of the decomposition. Our findings for
classical and exceptional algebras are summarized in Table \ref{tab:classical} . We subsequently show that a neural network
can also predict with high accuracy, the presence or absence of a given irreducible representation of a maximal sub-algebra within an irreducible representation of a parent algebra. The neural network is capable of predicting, for example, 
the presence of bi-fundamentals in $SU(3)\times SU(2)$ for a given representation of $SU(5)$ to an 
accuracy of 88\% and a confidence of 0.735.

We remark that our classification problems were also addressed with various standard classifiers, such as Naive Bayes, nearest neighbours and support vector machines.
We found that the NN with the architecture shown below in Fig. \ref{f:NNclass} significantly out-performed them. 
For example, using Logistic Regression for the analysis of Table \ref{tab:keras} for the $A_m$ algebra yields
a test accuracy of 0.823 and a confidence of 0.64. The results from support
vector machines are similar. This is in line with previous
observations where
NNs with similar architures 
perform well for a variety of problems, such as the computation of topological invariants of manifolds \cite{He:2017aed,Deen:2020dlf}, and finite graph invariants \cite{He:2020fdg}.

\section{Tensor Products and Branching Rules learnt by a Neural Network.}
\subsection{Tensor Products}\label{subsec:Tensor}
\subsubsection{Predicting the length of generic tensor decompositions}
Let us begin with a simple ML experiment.
One of the most important computations for Lie groups/algebras is the decomposition of the tensor product of two representations into a direct sum of irreducible representations for a given group $G$:
$
R_1 \otimes R_2 = \bigoplus\limits_{r \in \mbox{ irreps}} a_r R_r
$,
where $a_r \in \IZ_{\geq 0}$ are the multiplicity factors.
To be concrete, let us first consider $A_m = SU(m+1)$. Every irreducible representation (``irrep") of $A_m$ is specified by a highest-weight vector $\vec{v}$, which is a {rank} $m$ vector of non-negative integer components.
Throughout this letter, we will use
\begin{equation}
 \vec{v} \equiv (v_1, \dots, v_r)   
\end{equation}
 to denote the weight vector for a Lie algebra of rank $r$.
 When the context is clear, an integer with the vector over-script is understood to be a vector of the same integer entry, e.g., $\vec{4} = (4,4,\ldots,4)$. 
 
As the entries of $\vec{v}$ increase in magnitude, the dimension of the corresponding irrep $R_{\vec{v}}$ can grow dramatically.
For instance, for $A_3 = SU(4)$, $\dim R_{\vec{v} = (a,b,c)} = \frac{1}{12} (a+1) (b+1) (c+1) (a+b+2) (b+c+2) (a+b+c+3)$.
This makes the task of identifying the precise irreps contained in a tensor decomposition rather laborious. 

\begin{figure}[!h!t!b]
\centerline{
\includegraphics[trim=0mm 0mm 0mm 0mm, clip, width=3in]{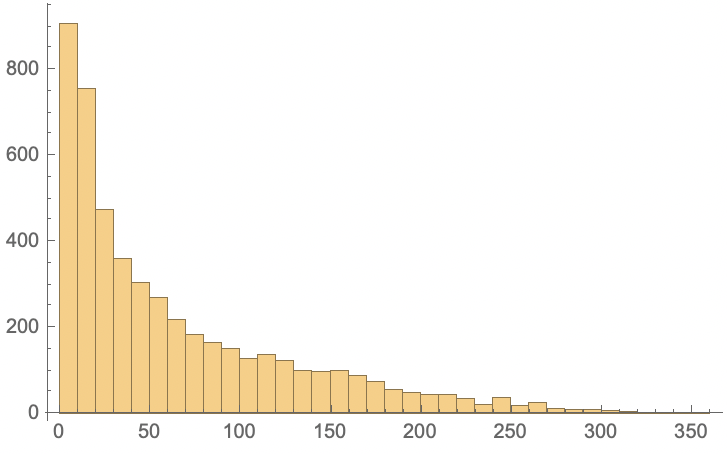}
}
\caption{{\sf {\small
Distribution of the number of distinct irreps in the tensor decomposition of $R_{\vec{v}_1} \otimes R_{\vec{v}_2}$ for $A_m$ with randomized $1 \leq m \leq 8$ and randomized ternary weight vectors $\vec{v}_{1,2}$. The horizontal axis denotes the length of $R_{\vec{v}_1} \otimes R_{\vec{v}_2}$ and the vertical axis denotes the corresponding frequency.
}}
\label{f:histv1v2len}}
\end{figure}

We start with two weight-vectors $\vec{v}_1,\vec{v}_2$. Their rank $m$ is chosen randomly from \{1,2,...,8\}. Then, we randomly generate a pair quinary vectors $\vec{v}_1,\vec{v}_2$ of rank $m$, and compute their tensor decomposition into irreducible representations:
\begin{equation}
R_{\vec{v_1}} \otimes R_{\vec{v_2}} = \bigoplus\limits_{r \ \in \mbox{ irreps}} a_r R_r \ .
\label{decomp}
\end{equation}  
This computation, although algorithmic, is non-trivial.
Even the relatively simple question of {\it how many distinct irreps, along with their multiplicities, are there on the RHS} or what we call the {\em length} of a given tensor decomposition, is not immediately obvious just by looking at the the vectors $\vec{v}_1$ and $\vec{v}_2$.
For example,
\begin{align}
\nonumber
R_{[0,1]} \otimes R_{[2,1]} &=  {\bf 8} \oplus {\bf 10} \oplus {\bf 27} \ ;\\
R_{[1,0,1]} \otimes R_{[0,2,0]} &= {\bf \overline{45}} \oplus {\bf 20'} \oplus {\bf 175}  \oplus {\bf 45} \oplus {\bf 15} \ .
\end{align}
It is difficult to see a priori that one decomposition would be of length 3 while the other would be of length 5; and one needs to actually compute the respective tensor decomposition to know the answer.
It took several hours using LieART to perform five thousands decompositions\footnote{
Care must be taken to find five thousands distinct pairs $(\vec{v_1},\vec{v_2})$ amidst the randomizations so as not to bias the input.
}.
To get an idea of their distributions, we show the histogram of the length: indeed there is a huge variation from 1 to over 350. 
A significant improvement in the running time (from hours to a few minutes) can be attained by capping off the maximum dimension of the irreps (say to 10,000). 
The distribution of the lengths of the decompositions vs frequency histogram, is depicted in figure \ref{f:histv1v2len}. 

\begin{figure*}
\includegraphics[]{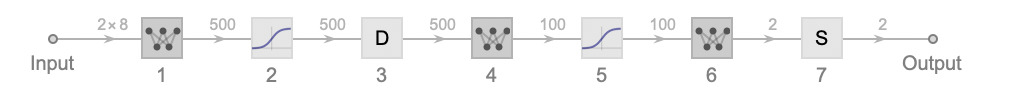}
\caption{The neural network architecture. $S$ is the softmax activation
and $D$ is a dropout layer with probability $0.2$. The hidden-layer neurons $(1,2)$ and $(4,5)$ are fully connected and sigmoid activated.}\label{f:NNclass}
\end{figure*}

Let us next consider a simple binary classification problem using the data generated by LieART: can ML distinguish tensor decompositions of length $\geq 70$ and of length $< 70$? The length $70$ is chosen since it splits the data rather evenly into around five thousands each.
To uniformize the input vectors, for the rank $m<8$, we also pad both $\vec{v}_{1,2}$ to the right with $-1$ (a meaningless number in this context) and stack them on top of each other.
Thus, our input is a $2 \times 8$ matrix with integer entries for $1\le m \le 8$. This step is essential for using a single NN for learning data for Lie algebras of varying ranks (it is for $A_m,1<m<8$ here). For the
majority of our experiments, we use a feed-forward neural network classifier built in $\mathtt{Mathematica}$
with the architecture shown in
Figure \ref{f:NNclass}.
We also reproduced these results with a similar 2-layer architecture on $\mathtt{Keras}$ \cite{keras},
with $\mathtt{selu}$ activated neurons to obtain similar accuracy and confidence. 
Finally, we need to ensure that the last softmax is rounded to 0 or 1 according to our binary categories.

\begin{figure}[t!!!]
\begin{subfigure}{7.5cm}
\includegraphics[trim=0mm 0mm 0mm 0mm, clip, width=7.5cm]{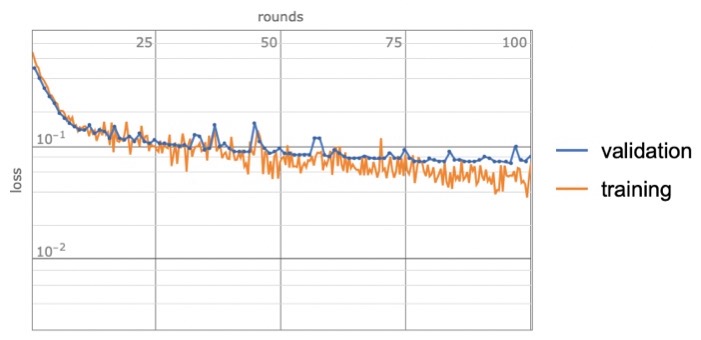}
\end{subfigure}
\begin{subfigure}{7.5cm}
\includegraphics[trim=0mm 0mm 0mm 0mm, clip, width=7.5cm]{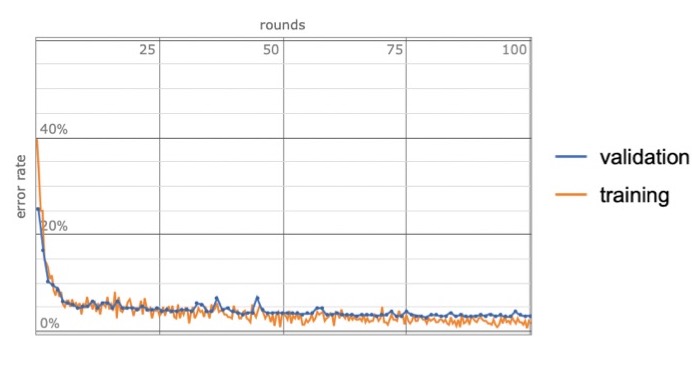}
\end{subfigure}
\caption{{\sf {\small
Loss-function (top), and error-rate (below) for training and validation for $A_m$ are plotted against number of epochs.
}}
}\label{f:Loss-n-Error-rate-exp1An}
\end{figure}

The results of our training and learning for 
$A_m$ are depicted in figure \ref{f:Loss-n-Error-rate-exp1An}.
The data was partitioned into 64\% training, 16\% validation, and 20\% test splits. The network was trained on the training and validation sets
and the test set was used purely for evaluating the trained network.
The plots show a steady lowering in both the error-rate and loss-function as we increase the number of rounds of training and validation.
We achieved accuracy 0.969, confidence 0.930, 5\% error rate, and 0.1 loss function within one minute by training for 100 epochs using learning rate $10^{-3}$, ADAM optimizer; which is excellent indeed.
Throughout this letter, we will use ``accuracy'' to mean percentage agreement of predicted and actual values. In addition, in discrete classification problems it is also important to have a measure of ``confidence'' so that false positives/negatives can be noted. A widely used one is the so-called Matthews' Phi-coefficient $\phi$  (essentially a signed square-root of the chi-squared of the contingency table) \cite{phi}, which is $\lesssim 1$ for predictions
with good confidence.
 

The above experiment was also carried out with other classical, as well as exceptional Lie algebras with comparable success. 
The results are given in table~\ref{tab:classical}. We generated the same data size as in the $A_m$ case, i.e. 5000, and used the same cap on the maximum dimension of the irreps (10,000). In contrast, though the dimension cap for exceptional groups
was set to 120,000, it yielded far fewer data points.
The lengths we split the data-sets on were chosen to 
generate a balanced data-set in each case. 
The accuracy of ML prediction was above .95 for each of these cases.
\begin{table}[ht]
\centering
\begin{tabular}{|c|c|c|c|c|c|}
    \hline
    Group & Data Size & Splitting Length & Accuracy & Confidence\\
    \hline
    $A_m$ & 5000 & 70 &  0.969 & 0.930\\
    \hline
    $B_m$ & 5000 & 40 &0.959 & 0.878\\
    \hline 
    $C_m$ & 5000 & 40 &  0.969 & 0.921\\
    \hline 
    $D_m$ & 5000 & 35 &  0.965 & 0.908\\
    \hline
    $G_2$ & 1275 & 110 &  0.946 & 0.891\\
    \hline
    $E_6$ & 903 & 30 &  0.898 & 0.795\\
    \hline
\end{tabular}
\caption{
The binary classification of product decomposition lengths. The splitting lengths yield a balanced dataset.}\label{tab:classical}
\end{table}
The relatively lower accuracy for $E_6$ is caused by the low number of points available at low dimensions due to its relatively high rank: 903 data points below dimension of 120,000. Raising the dimension cap would improve the machine-learning, bringing it up to par with others, however the corresponding data generation using LieART would take days.

We also note that partitioning the data-sets at the `midpoint' to generate
balanced
data-sets as we have done above is by no means necessary. As an example, we 
explored this classification 
problem for the $A_m$ algebras but now organizing the data
into partitions of varying lengths, 
\textit{viz.} 20/80 through to 80/20. Here by a partition of length 20/80
we mean that a `cutoff' decomposition length was chosen such that 20\% of the
decomposition lengths in the dataset are below this length, i.e. are denoted by the target variable $\mathrm{Y}=0$ and the remaining
80\% are above, and hence denoted by $\mathrm{Y}=1$.
In every case the
Matthews' Phi-coefficient remains close to 1. In particular, for the 
20/80 and 80/20 partition it is 0.98.

We can take this experiment one step further and train the neural net on low dimensional tensor decomposition data, then test its performance on higher dimensional cases. If successful this would immensely reduce the computation time. For example, obtaining the length of decomposition for two $A_6$ weight vectors $\vec{v}_1=\vec{v}_2=(2,2,2,2,2,2)$ by brute force takes over 15 minutes on LieART while machine learning should estimate the length in a matter of seconds.

\begin{figure}[!h!t!b]
\begin{subfigure}{6.5cm}
\includegraphics[trim=0mm 0mm 0mm 0mm, clip, width=6.5cm]{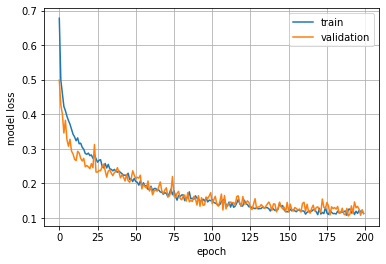}
\end{subfigure}
\begin{subfigure}{6.5cm}
\includegraphics[trim=0mm 0mm 0mm 0mm, clip, width=6.5cm]{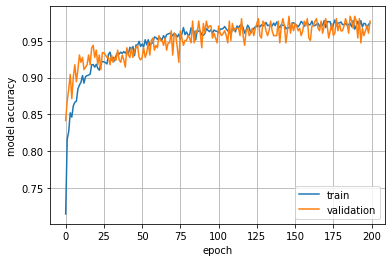}
\end{subfigure}
\caption{{\sf {\small
Loss-function (top), and model accuracy (below) for training on $A_m$ irreps, plotted against number of epochs.
}}
}\label{f:keras}
\end{figure}

We retrained the NN in figure~\ref{f:NNclass} 
on the same data for the classical
and $G_2$ algebras generated by LieART for the previous experiment.
However, the training set is now restricted to have both input weight vectors of dimension less than certain cut-off value, here taken to be 2,000. The
\textit{trained} neural network was subsequently evaluated on the test 
dataset consisting of input weight vectors of dimension ranging between 2,000 and 10,000.
Our results are
presented in Table \ref{tab:keras} and Figure \ref{f:keras}. 
\begin{table}[h]\label{Table4}
\centering
\begin{adjustbox}{width=0.45\textwidth}
\begin{tabular}{|c|c|c|c|c|c|}
    \hline
    Group  & Train/Val Accuracy & Test Accuracy & Confidence\\
    \hline
    $A_m$ & 0.974/0.957 & 0.961 &0.907\\
    \hline
    $B_m$  & 0.972/0.963 & 0.957 &0.845\\
    \hline 
    $C_m$  & 0.969/0.970 & 0.892 &0.792\\
    \hline 
    $D_m$  & 0.971/0.940 & 0.956 &0.817\\
    \hline
    $G_2$  & 0.969/0.963 & 0.968 &0.922\\
    \hline
    $E_6$  & 0.963/0.947 & 0.875 &0.751\\
    \hline
\end{tabular}
\end{adjustbox}
\caption{
Training on low dimensional irreps, and testing on high dimensional ones. }\label{tab:keras}
\end{table}
\subsubsection{Beyond Binary Classification}
\begin{figure}[h]
\centerline{
\includegraphics[trim=10mm 0mm 0mm 0mm, clip, width=7.5cm]{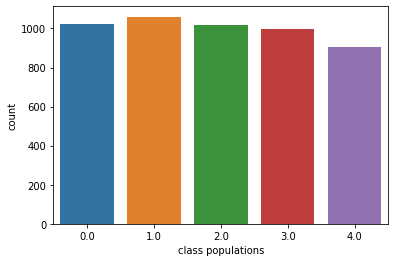}
}
\caption{{\sf {\small
Class Populations of length bins for $A_m$ irreps.
}}
\label{f:hist}}
\end{figure}

\begin{figure}[]
\begin{subfigure}{8cm}
\includegraphics[trim=0mm 0mm 0mm 0mm, clip, width=7.cm]{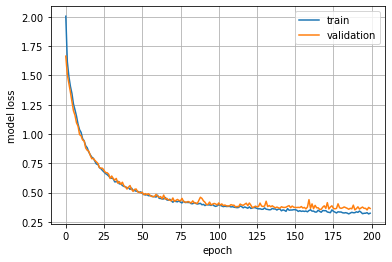}
\end{subfigure}
\begin{subfigure}{8cm}
\includegraphics[trim=0mm 0mm 0mm 0mm, clip, width=7.cm]{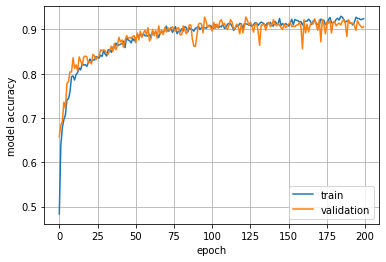}
\end{subfigure}
\caption{{\sf {\small
Training curves for the quinary classification problem for $A_m$, plotted against number of epochs.
}}
}\label{f:5bintraining}
\end{figure}

We now move beyond the simpler binary classification experiments done 
previously to a multi-class classification task, with the aim of predicting a
range for the length of the product decomposition as opposed to the over/under
estimates obtained above. For definiteness, let us take the $A_m$ data and
classify it into five classes, depending on whether the length of the product
decomposition lies in the ranges 0 to 10, 10 to 25, 25 to 55, 55 to 115 and
greater than 115. Figure \ref{f:hist} shows a histogram with the class populations, and the training curves are displayed in Figure
\ref{f:5bintraining}. 
The neural network reaches a $\phi$-coefficient of 0.917
on the test set, and the confusion matrix is given by
\begin{equation}
\begin{pmatrix}
96 &  3 &  0 &  0 & 0\\
   1 & 92 &  3 & 0  & 0\\
   0  & 2 & 109 &4  & 0\\
   0  & 0 &  7 & 86& 7\\
   0  & 0 &  0 & 6  & 84
\end{pmatrix}
\end{equation}

\subsection{Branching Rules}\label{subsection:Branching}
\begin{figure}[h!!!]
\centerline{
\includegraphics[trim=10mm 0mm 0mm 0mm, clip, width=3in]{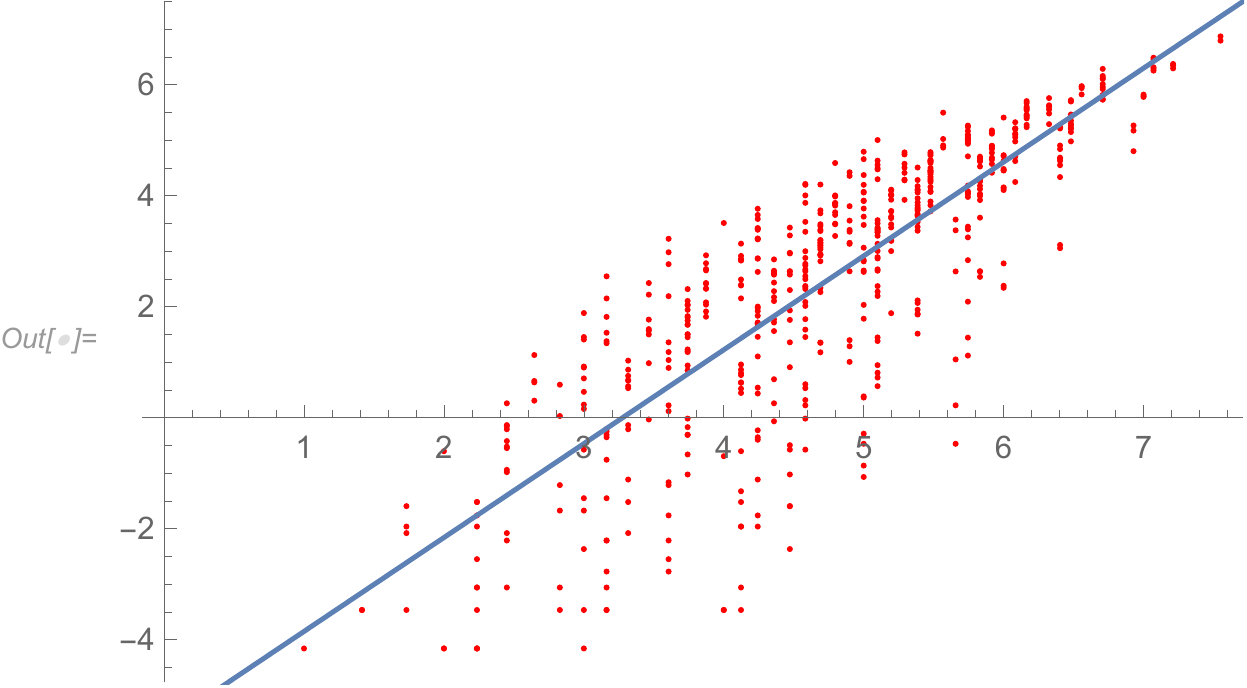}
}
\caption{{\sf {\small
Plot and fit of log of the time in seconds of the branching of irreps of $SU(5)$ versus the length of the weight-vector.
}}
\label{f:timingBranchingSU5}}
\end{figure}

The next task on which we train our neural network of Figure
\ref{f:NNclass} is to learn about the branching rules for Lie algebras.
Suppose we take a  weight-vector of $SU(5)$, and restrict its entries from 0 to 4 (i.e., as quinary 4-vectors). 
Even though this may look rather harmless, the dimension of the corresponding irrep ranges from 1 for $\vec{0}$, to 9765625 for $\vec{4}$.
When we decomposed these irreps of $SU(5)$ to those of its maximal sub-algebra $SU(3) \times SU(2) \times U(1)$, and found their explicit branching products, the time taken on LieART was easily seen to be exponential\footnote{ Notice that as LieART is only capable of generating branching rule data for maximal subgroups, here we will focus on this simplest set of branching training data to illustrate the capability of neural network.}.
In figure \ref{f:timingBranchingSU5}, we plot the log of the time taken in seconds, versus the length of the weight-vector.
The best fit is the line $-5.54361 + 1.69186 x$.
By extrapolation, the single irrep of $SU(5)$ corresponding to weight vector $\overrightarrow{10}$ would take over 20 years just to compute its branching products into $SU(3)\times SU(2) \times U(1)$.

In the rest of this section, we shall show the efficacy of using ML to predict presence/absence of any given representation of the maximal sub-algebra in a given irrep of $SU(5)$ and $G_2$ algebras. For concreteness, we look for bi-fundamental representations of $SU(3)\times SU(2)$ (with arbitrary values of $U(1)$ charges) in any given $SU(5)$ irrep. 
In the $G_2$ case, we restrict ourselves to bi-fundamental representation of $SU(2)\times SU(2)$ maximal sub-algebra.

For the $SU(5)$ branching, we use first 800 irreps of smallest dimension as
input vectors with binary output depending on presence/absence of a bi-fundamental rep of $SU(3)\times SU(2)$. The data was split into train/validation/test sets in the ratio 80/10/10. The neural network reached
a test accuracy of 0.899 and a confidence of 0.813. The next best results
were arrived at by a support vector classifier which reached an test accuracy
of 0.838 and confidence of 0.677.

For the $G_2$ branching, we used 400 weight input vectors with dimensions below 4.7 million. Analogous to the $SU(5)$ case, the output was binary, depending on presence/absence of a bi-fundamental rep of $SU(2)\times SU(2)$. 
All classifiers, neural nets and otherwise, performed at the level of blind 
guessing in this case, which is possibly due to the relatively fewer input 
data as well as smaller number of features in the data. 

\section{Outlook}
Given the ubiquity of Lie algebras and groups in physics, let us end this letter with some comments about the vast possibilities in applications to physics of our results, exemplifying with two which immediately come to mind.

In scattering processes, given a pair of incoming particles transforming under the irreps of certain global symmetry group, the outgoing particles can be classified via their tensor decompositions.
The tensor decomposition prediction and extrapolation results in section \ref{subsec:Tensor} thus allow us to efficiently estimate the number of distinct outgoing particles. It would also be exciting to see if the NN upper bound estimate of the length of a given decomposition can help LieART package to work out its explicit terms within significantly shorter period.

Our choice of studying the branching of $SU(5)$ into its maximal subgroup $SU(3)\times SU(2)\times U(1)$ in section \ref{subsection:Branching} was phenomenologically motivated. This hopefully can lead to an useful algorithm for testing whether a field transforming under $SU(5)$ GUT gauge group can yield descendants transforming under standard model gauge groups upon spontaneous symmetry breaking. We hope this will be useful for particle physics model building purposes.

\acknowledgments
 HYC is supported in part by Ministry of Science and Technology (MOST) through the grant
108-2112-M-002 -004 -. 
YHH is indebted to STFC UK, for grant ST/J00037X/1.
SL’s work was supported by the Simons Foundation grant 488637 (Simons Collaboration on the Non-perturbative bootstrap) and the project CERN/FIS-PAR/0019/2017. Centro
de Fisica do Porto is partially funded by the Foundation for Science and Technology of Portugal (FCT) under the grant UID-04650-FCUP. SM's work is supported by SMCSE Doctoral Studentship at City, University of London.


\end{document}